\documentclass[12pt,preprint]{aastex}

\usepackage{graphicx,amsmath}

\slugcomment{accepted for publication in ApJL}

\shorttitle{Apodized Pupil Lyot Coronagraphs}
\shortauthors{R. Soummer}

\begin{document}

\title{Apodized Pupil Lyot Coronagraphs for arbitrary telescope apertures}

\author{R\'emi Soummer. \altaffilmark{1}}
\affil{Space Telescope Science Institute, 3700 San Martin Drive, Baltimore MD21218, USA}
\email{soummer@stsci.edu}

\altaffiltext{1}{Michelson Fellow}

\begin{abstract}
In the context of high dynamic range imaging, this study presents a breakthrough for the understanding of Apodized Pupil Lyot Coronagraphs, making them available for arbitrary aperture shapes. 
These new solutions find immediate application in current, ground-based coronagraphic studies (Gemini, VLT) and in existing instruments (AEOS Lyot Project). They also offer the possiblity of a search for an on-axis design for TPF. The unobstructed aperture case has already been solved by Aime et al. (2002) and Soummer et al. (2003).
Analytical solutions with identical properties exist in the general case and, in particular, for centrally obscured apertures. 
Chromatic effects can be mitigated with a numerical optimization.
The combination of analytical and numerical solutions enables the study of the complete parameter space  (central obstruction, apodization throughput, mask size, bandwidth, and Lyot stop size).
 \end{abstract}

\keywords{techniques: high angular resolution, instrumentation: adaptive optics}

\section{Introduction}
High dynamic range imaging is a very exciting and promising new field of astronomy and astrobiology. Its principal goal is the  detection and characterization of exoplanets. It also enables the study of faint sub-stellar companions, and circumstellar debris and
protoplanetary disks, e.g.  \citet{AS03,AS04b}.

The detection of faint sources around stars is limited by several 
sources of noise, mainly due to the diffracted starlight.
This light has to be removed using coronagraphic or nulling 
instruments. In reality, such instruments are sensitive 
to wavefront errors and temporal instabilities,
but their performance can be predicted from the results in the perfect case
using knowledge of the wavefront errors. This can be approached
with expansions or statistical methods 
\citep{SLH02, PSM03,AS04}.

The starlight leakage through a coronagraph consists of residual 
speckles \citep{SKM01}, and techniques for their reduction and calibration 
needs to be developed \citep{SF02,MRD04, Guy04}.
If the system is limited by wavefront errors, which is the case on most 
existing systems \citep{BRB04}, it is not worth working to improve their static extinction \citep{AS04}.
With additional speckle reduction techniques, this limit might be pushed
back.

The most exciting current project, the Terrestrial Planet Finder (TPF) requires a 
magnitude difference of the order of $\Delta m = 25$ at $0.1\arcsec$ in 
the visible, which is far beyond what is currently achievable.
While early lab demonstrators are necessary along this path, actual precursor science must 
also pave the way with meaningful science results \citep{SWG04}.
Unless a TPF-precursor is funded, for example through a NASA 
Discovery class mission, most of the coronagraphic precursor science 
will be done on the ground by the next generations of Extreme Adaptive 
Optics  (ExAO) coronagraphs followed by Extremely Large 
Telescopes. Several ground-based observatories are currently 
conducting preliminary studies for such instruments (ESO/VLT-PF, 
Gemini, Subaru, Palomar).

The near-IR Lyot Project coronagraph on AEOS \citep{Lyot04} 
uses the highest order AO system currently available to the civilian astronomical 
community.  It operates in the regime where 
improved coronagraphs start to produce greater science returns.
It is clear that the next generation of ExAO coronagraphs under study will need a new generation 
of coronagraph designs.

Ground-based coronagraphs have to deal with centrally obscured telescope geometries, at least for
the current generation of instruments. It is therefore essential to develop some theoretical insight and 
understanding on how a coronagraph can work under such constraints.
This is also relevant to TPF, where an off-axis design --- the only existing
design that meets TPF goals ---  has an  extremely 
tight tolerancing budget. 
If a coronagraph were able to work at the required contrast level 
with a central obstruction, the error budget and complexity
of the TPF design could be relaxed. Spin-offs of such a design, 
such as the possibility of additional wide field instrumentation,
could also make TPF more interesting to the general astronomical community.

The need to apodize unobstructed apertures has been demonstrated for Roddier phase mask \citep{RR97} and Lyot coronagraphy.

and particular solutions have been found \citep{ASF01,ASF02,SAF03}.
However, even without wavefront errors, these apodizations \citep{ASF01,ASF02,SAF03} cannot be 
applied directly to centrally obscured apertures and in wide bandpasses without considerable 
loss of efficiency \citep{SAF03b}.
This effect has been also shown by Abe (2003, Private communication), using empirical apodizations to improve performance.
This letter generalizes unobstructed pupil results to arbitrary aperture shapes;
it presents a breakthrough in our current understanding of Lyot coronagraphy,  
and offers hope for further improvements.
Apodized Pupil Lyot Coronagraphs (APLC) present the advantage of simplicity and lower sensitivity to errors compared to other techniques insensitive to central obstruction \citep{SDA03,GR96}.

\section{General problem of coronagraphy with arbitrary aperture shapes}
Following the notation of \citet{ASF02} and \citet{SAF03}, we briefly recall the general formalism of coronagraphy with apodized pupils.
The telescope aperture function with the position vector $\textbf{r}$ is denoted by $P(\textbf{r})$, and $\Phi(\textbf{r})$ is the apodizer transmission (1 without apodization). At the entrance aperture, the wavefront amplitude is:
\begin{equation} \label{Eq1}
\Psi_A(\textbf{r})=P(\textbf{r})\, \Phi(\textbf{r}).
\end{equation}
A mask of transmission $1-\epsilon M(\textbf{r})$ is placed in the focal plane. $M$ is a function that describes the mask shape, equal to 1 inside the coronagraphic mask and 0 outside ($\epsilon=1$ for Lyot and  $\epsilon=2$ for Roddier).

At the Lyot stop plane, the wavefront is then:
\begin{equation}\label{Eq2}
{\Psi_C}(\textbf{r}) =
 P(\textbf{r})\, \Phi(\textbf{r}) - \frac{\epsilon}{\lambda^2f^2}   \,\int_{P}{{\Phi}}(\textbf{u})\, \widehat{M}\left(\frac{\textbf{r}-\textbf{u}}{\lambda f}\right)\, \mathrm{d}\textbf{u},
  \end{equation}
  where  $ \widehat{  } $  is the Fourier Transform. This general relation is valid for any arbitrary aperture shape (rectangular, circular or elliptical, with or without central obstruction and secondary mirror supports), and any mask shape. 

\begin{figure}[ht]
\plotone{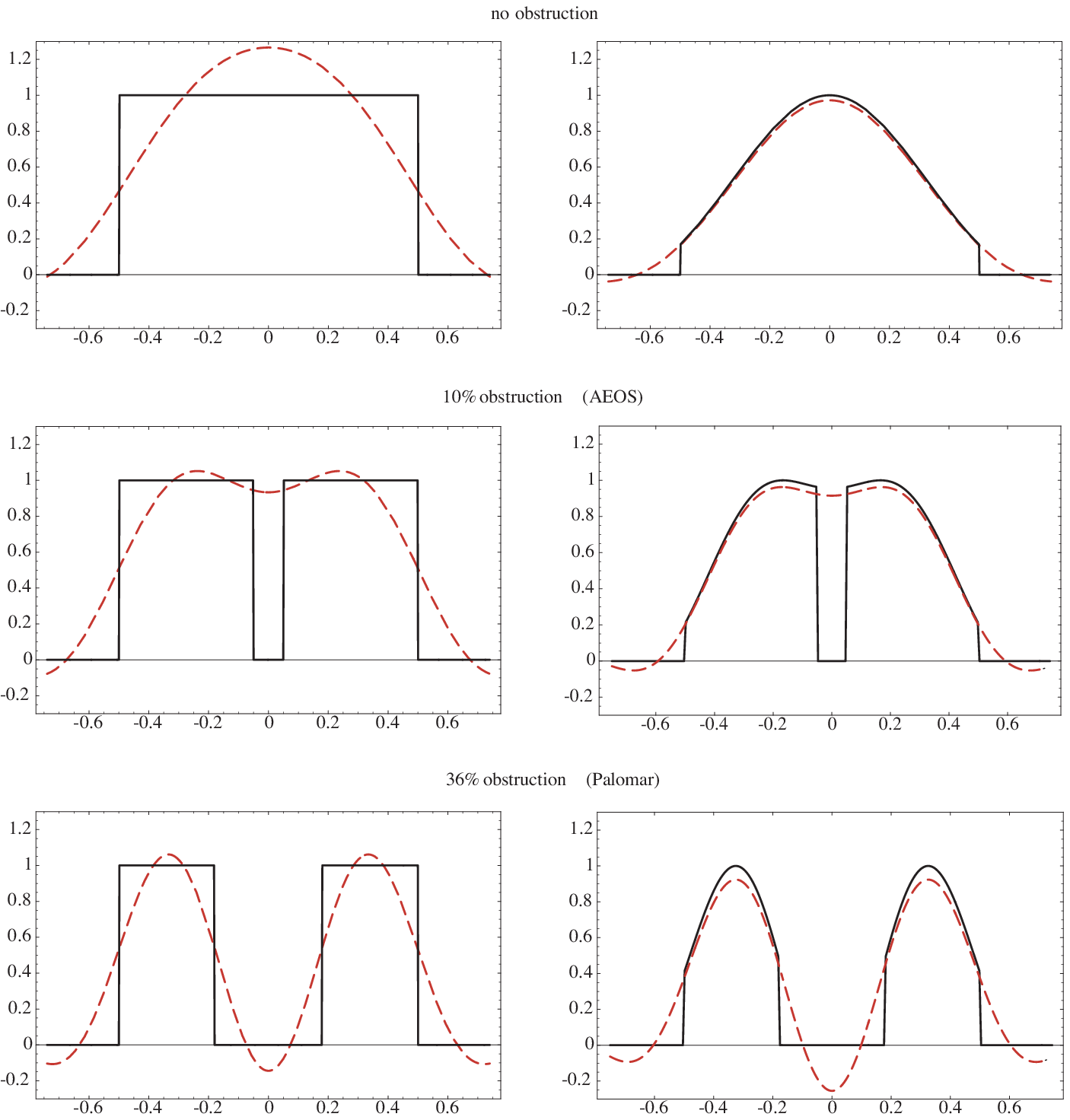} 
 \caption{Transverse cuts of the two wavefronts that subtract one another in the Lyot stop Plane (Eq.\ref{Eq2}): the pupil wave (solid line) and the wave diffracted by the mask (dashed line), for a clear circular aperture (top), then for small and large obstructions. In the non-apodized case (left), the two waves clearly don't match each other and the subtraction is not efficient. With apodized apertures (right), the match is much better and the two curves are proportional inside the aperture (Generalized prolate solutions). The mask sizes have been chosen for better visualization. } \label{f1}
\end{figure}

The formal problem is to find the mask $M(\textbf{r})$ and the apodizer $\Phi(\textbf{r})$ that provide the best cancellation possible inside the Lyot stop, identical to the entrance pupil in this case. 
The above partition of the wavefront (Eq.\ref{Eq2}) gives a physical understanding of coronagraphy as a destructive interference between the pupil wave ${\Psi_A}(\textbf{r}) $ and the wave diffracted by the mask (the convolution integral). A heuristic illustration is given in Fig.\ref{f1}, for a circular geometry.
In the unapodized case (left), the two wavefronts do not match each other and the subtraction leaves a large residual field.
The APLC solution is to smooth out the pupil transmission so that the wave diffracted by the mask matches the pupil wave. The improvement of the wave subtraction is obvious in the Fig.\ref{f1} (right).
An opposite approach is to use a clear aperture, and instead modify the shape of the wave diffracted by the mask to be a flat function in a reduced part of the aperture \citep{KT02}.

A perfect solution (total extinction) is obtained if the wave diffracted by the mask is equal to the initial pupil amplitude. 
In more mathematical terms, this means that $\Phi(\textbf{r})$ must be the eigenfunction of the convolution integral in Eq.\ref{Eq2}.
This problem is equivalent to the generalization of the uncertainty principle \citep{Pap68}, with the search of band-limited functions having the maximum encircled energy within a given spatial domain.
In coronagraphic terms, the solution is the pupil apodization which produces the most concentrated star light behind a given focal plane mask (and thus blocked).
This formal problem has been solved analytically for rectangular and circular unobstructed apertures \citep{ASF01,ASF02,SAF03} involving linear and circular prolate functions \citep{SP61,Fri71,Pap81}.
In the general case, formal solutions exist \citep{Sle64,Sle65,Pap68} and are the eigenfunction of the 2-D integral equation:
\begin{equation}\label{Eq3}
\int_{P}{{\Phi}}(\textbf{t})\, \widehat{M}(\textbf{r}-\textbf{t})\, \mathrm{d}\textbf{t}=\Lambda\, \Phi(\textbf{r}).
\end{equation}
The solution  $\Phi(\textbf{r})$ corresponds to the largest eigenvalue $\Lambda_0$, and to the maximum encircled energy behind the focal plane mask $e=\Lambda_0$.
The expression of the residual wavefront inside the Lyot stop for circular and rectangular geometries is the same for the general case:
\begin{equation}\label{Eq4}
{\Psi^+_C}(\textbf{r}) =
  {\Psi_A}(\textbf{r})(1 - \epsilon \, \Lambda_0),
  \end{equation}
(the $^+$ signifies this is just after the Lyot stop).
As detailed in previous work, the wavefront is therefore simply proportional to the initial apodization function, with a constant of proportionality $(1-\epsilon\,\Lambda_0)$. 
This creates the possibility for multiple stage coronagraphs \citep{ASF02,AS04c}, which could help TPF reach its required extinction. 

A numerical algorithm, proposed by \citet{GR00} for the phase mask, can be used to generate the prolate apodizations for the APLC \citep{SAF03b}. 
Starting with an initial guess for the apodizer, the principle is to subtract the residual Lyot stop wave amplitude from the aperture wave amplitude in an iterative loop.
\citet{GR02} used this algorithm to show that the Roddier phase mask can be used with arbitrary apertures. This algorithm works for both the Lyot as well as the general case.
A formal proof of its convergence in the general case, using mathematical properties of the generalized prolate functions, along with more detail on the properties of the solutions for coronagraphy, will be given in a future work. 

\section{Specific solutions for circular apertures with central obstructions}
For circular aperture telescopes, Eq.\ref{Eq2} can be simplified using the properties of Hankel Transforms, and using a kernel $\mathrm{K_0}(\xi ,r)$ which is a function of the focal plane mask size \citep{Fri71,SAF03}. The equation can be generalized to:
\begin{equation}\label{Eq5}
{\Psi_C}(r) =   {\Psi_A} (r) - \epsilon \,{\left( 2\,\pi  \right) }^2 \int_{D_2/2}^{D_1/2}  \xi \, {\Phi} (\xi)\,{\mathrm{K}_0}(\xi ,r)\,d\xi,
\end{equation}
where $D_1$ is the primary aperture and $D_2$ is the secondary aperture. 
Solutions exist as a subset of the general problem and involve a set of new functions -annular prolate functions- with identical properties to circular prolate functions.  

As illustrated in Fig.\ref{f1}, apodizers can have different shapes depending on the size of the central obstruction, with important consequences for relative throughput and performance. 
A small obstruction produces an asymmetrical apodizer shape; the larger obstruction has a more symmetrical apodizer. 
This effect depends both on mask size and central obstruction: the term $\widehat{M}$ in the convolution of Eq.\ref{Eq2}  extends over a size of the order of $D_1/\alpha$ where $\alpha$ is the mask size in units of $\lambda f /D_1$. It is therefore relevant to compare $D_1/\alpha$ and $D_2$. 

If the extent of $\widehat{M}$ is much larger than the central obstruction (small mask and/or small obstruction), the convolution product smooths out the central obstruction, and the result is an apodizer with maximum transmission at or near the central obstruction (Fig.\ref{f1}, center). The resultant apodizer is very similar to that of the 'classical' unobstructed case. 

\begin{figure}[ht]
\epsscale{0.85}
\plotone{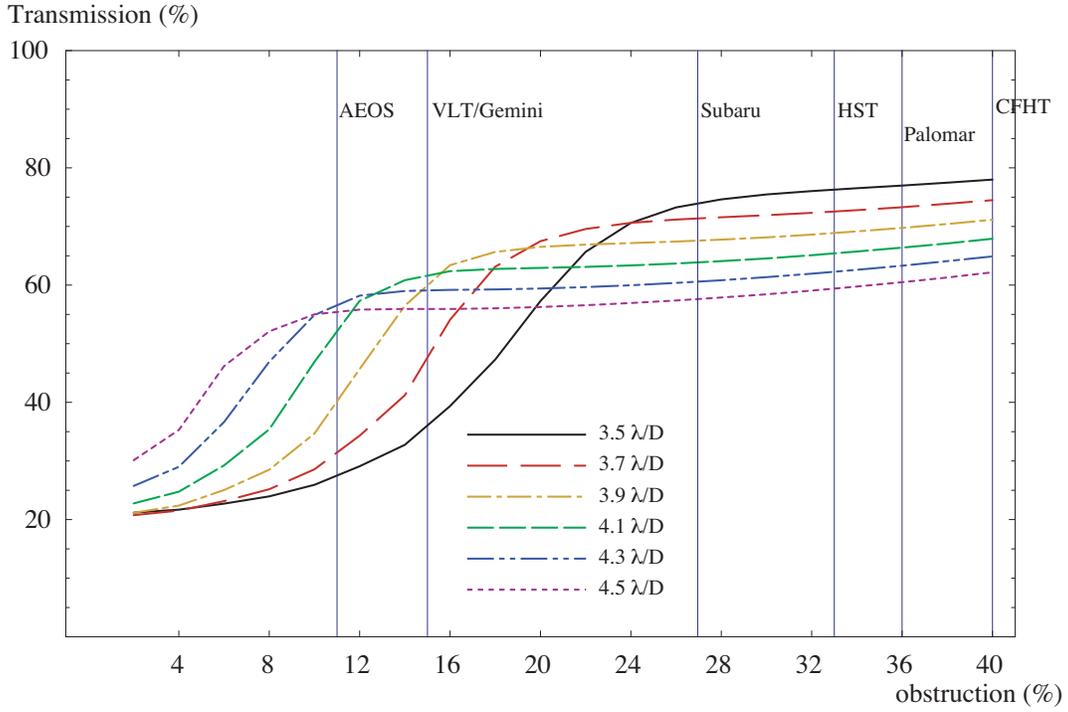}
\vspace{1in}
  \caption{Throughput of the apodizer solution to the eigenvalue problem as a function of the central obstruction, and for several mask sizes. Vertical lines indicate the geometry of existing telescopes. Apodizers throughput for large masks saturate for smaller apertures to a lower value of throughput.}
  \label{f2}
\end{figure}

\begin{figure}[ht]
\epsscale{0.80}
\plotone{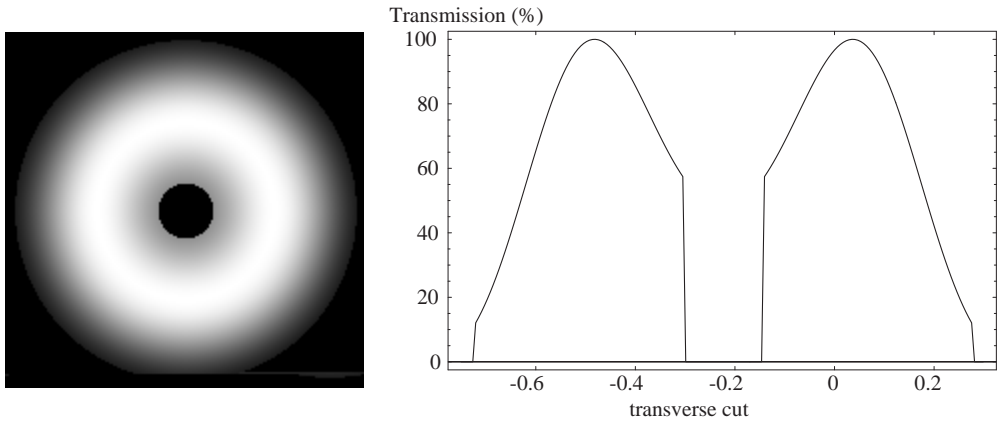} 
\vspace{1in}
 \caption{Example of apodizer transmission for the geometry of Gemini/VLT. The minimum intensity transmission is $12\%$ at the edges and throughput is high: $63\%$. A classsical Lyot coronagraph with undersized Lyot stop, has a typical throughput of 40-60\% . The corresponding mask has a diameter of $4\lambda/D$. }
  \label{f3}
\end{figure}

If the extent of $\widehat{M}$ is smaller than the central obstruction (large mask and/or large obstruction), the convolution simply smooths the inner and outer part of the pupil. The result is a symmetrical, "bagel"-shaped apodizer. This regime provides a very high relative throughput and is favorable in terms of the residual light distribution, itself apodized. Throughputs are illustrated in Fig.\ref{f2} for a large range of central obstructions, and for mask sizes between $3.5\lambda/D$ and $4.5\lambda/D$. The throughput curves present a shoulder corresponding to the transition from the classical shape to the 'bagel' shape (Fig.\ref{f3}). There is also a gain in angular resolution in this regime due to the lighter apodization, especially near the outer-edges of the pupil. Empirical apodizations with similar shapes were proposed and manufactured by \citet{JR64}.

Generalized prolate solutions are monochromatic. However, APLCs are sensitive to the PSF size chromaticity \citep{SAF03b}. Changing wavelength is formally equivalent to changing mask sizes. 
Since the coronagraphic rejection is not symmetrical around the optimum value, there is a substantial gain in re-optimizing the mask size for broad bandpasses, with respect to the maximum encircled energy within the mask. 

\begin{figure}[ht]
\epsscale{0.75}
\plotone{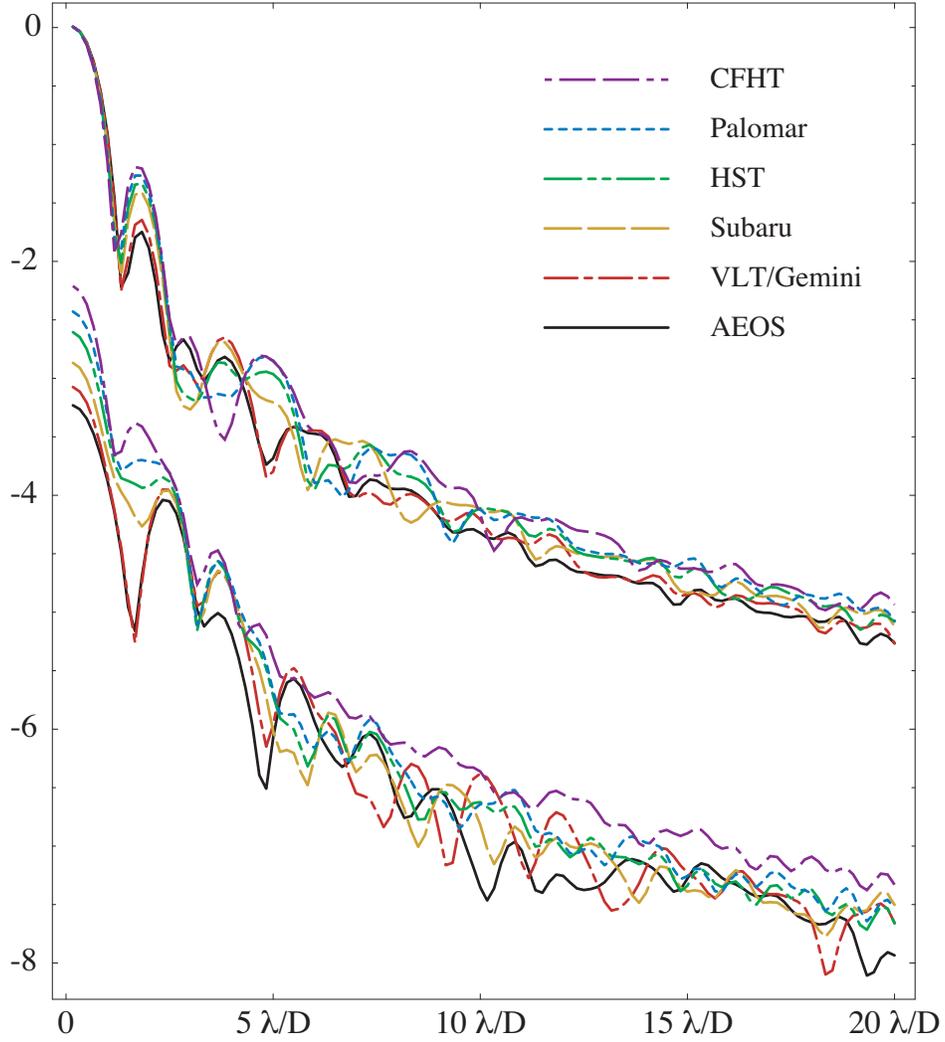} 
\vspace{1in}
 \caption{Log scale broadband coronagraphic PSFs  ($\Delta \lambda/\lambda=20\%$) corresponding to existing telescopes geometries (AEOS, Gemini, VLT, Palomar, HST, CFHT). The mask sizes are optimized for broadband and have a typical diameter of $4.2 \lambda/D$.  The peak value of the PSF is normalized to the one withouth focal plane mask. The top set of curves corresponds to the un-occulted PSFs. These results are obtained without wavefront errors.}
  \label{f4}
\end{figure}

Static coronagraphic PSFs are given in Fig.\ref{f4}, for a 20\% bandwidth and several telescope shapes. Table 1 gives the mask sizes values optimized for the bandpass.

\section{Conclusion}

This letter presents a breakthrough in the understanding of APLCs, proving the possibility of their use with arbitrary apertures, and sketching a method for their broadband optimization. 
This formalism enables the study of the entire parameter space (central obstruction, mask size, bandwidth, Lyot stop size). 
It is also possible to generate solutions taking into account secondary mirror support structures, increasing the complexity with azimuthal features in the apodizer.

This general formalism enables the study of apodized Shaped Pupils \citep{KVS03} hybrid designs, as well as multiple stage coronagraphy \citep{ASF01,AS04c}.

The static coronagraphic PSFs given here do not correspond directly to the dynamic range; a combination between the generalized prolate and the statistical approach of \citet{AS04} will be presented in a separate paper.


\acknowledgments
The author is supported by a Michelson Fellowship, under contract with the Jet Propulsion Laboratory funded by NASA. JPL is managed for NASA by the California Institute of Technology. 
The author would like to thank C. Aime, A. Sivaramakrishnan and R. Makidon for daily discussions and R. Frieden for comments and advice.

\clearpage

\clearpage

\begin{deluxetable}{lcccc}
\tabletypesize{\scriptsize}
\tablehead{
\colhead{Telescope} & \colhead{Obstr.} & \colhead{Apodizer} & \colhead{mask} & \colhead{PSF@$5\lambda/D$}}
\startdata
AEOS & 11\% &  55\% & 4.60 & $1.3\,10^{-6}$\\
VLT/Gemini & 16\% & 56 \% & 4.61 & $1.7\, 10^{-6}$\\
Subaru & 27\% & 58\% & 4.61 & $8.0  \, 10^{-7}$\\
HST & 33\% & 60\% & 4.51 & $1.0 \, 10^{-6}$\\
Palomar & 36\% & 62\% & 4.48 & $1.0  \, 10^{-6}$\\
CFHT & 40\% & 65\% & 4.48 & $2.3  \, 10^{-6}$\\
\enddata
\tablecomments{This table corresponds to the illustration of Fig.\ref{f4}}
\end{deluxetable}

\end{document}